# Viscous Strength of Water

**K.Y. Volokh**[1]

*Faculty of Civil and Environmental Engineering*

*Technion - Israel Institute of Technology, Haifa 32000, Israel*

## Abstract

In the laminar mode interactions among molecules generate friction between layers of water that slide with respect to each other. This friction triggers the shear stress, which is traditionally presumed to be linearly proportional to the velocity gradient. The proportionality coefficient characterizes the viscosity of water. Remarkably, the standard Navier-Stokes model surmises that materials never fail – the transition to turbulence can only be triggered by some kinematic instability of the flow. This premise is probably the reason why the Navier-Stokes theory fails to explain the so-called subcritical transition to turbulence with the help of the linear instability analysis. When linear instability analysis fails, nonlinear instability analysis is often resorted to, but, despite the occasional uses of this approach, it is intrinsically biased to require finite flow perturbations which do not necessarily exist.

In the present work we relax the traditional restriction on the perfectly intact material and introduce the parameter of *fluid viscous strength*, which enforces the breakdown of internal friction. We develop a generalized Navier-Stokes constitutive model which includes a material failure description, and use it to analyze the Couette flow between two parallel plates to find that the lateral infinitesimal perturbations can destabilize the laminar flow. Furthermore, we use the results of the recent experiments on the onset of turbulence in pipe flow to calibrate the viscous strength of water. Specifically, we find that the maximal shear stress that water can sustain in the laminar flow is about one Pascal. We note also that the introduction of the fluid strength suppresses pathological stress singularities typical of the traditional Navier-Stokes theory and uncovers new prospects in the explanation of the remarkable phenomenon of the delay of the transition to turbulence due to an addition of a small amount of long polymer molecules to water.

---

[1] E-mail: cvolokh@technion.ac.il

# 1. Introduction

There are at least *four* good reasons to encourage a generalization of the Navier-Stokes constitutive model for water.

***First***, material (namely water) never fails according to the traditional model. Let us assume that there are no finite disturbances in the shear flow between parallel plates. It is implied that any magnitude of shear stress can be reached by an appropriate acceleration of a plate. The stress is not bounded according to the traditional Navier-Stokes model! Our experience, however, teaches us that *all* materials fail and the maximal achievable stress is always bounded. The traditional model cannot amend this gap.

***Second***, experiments (Mullin and Kreswell, 2005; Barkley and Tuckerman, 2005; Prigent and Dauchot, 2005) show the transition to turbulence can start at Reynolds numbers lower than predicted by linear instability analysis – the subcritical transition to turbulence. As mentioned, nonlinear instability analysis is often resorted to in order to bridge this gap, but it requires *finite* flow perturbations which don't necessarily exist. Thus, the approach of nonlinear instability analysis is generally open to criticism from the physical standpoint. Infinitesimal (molecular) perturbations always exist so one should expect the linear instability analysis to catch the onset of the transition to turbulence.

***Third***, the classical Navier-Stokes model implies the existence of stress singularities, which are not real physical phenomenon. A physically sound theory should suppress any such singularities.

***Fourth***, the remarkable phenomenon of the delay of turbulence is observed when a small amount of polymer molecules is dissolved in water (Tanner and Walters, 1998). The traditional Navier-Stokes model is unable to account for this phenomenon in principle because the only material constant – the viscosity coefficient – is not affected by the small amount of added polymer. Clearly, more material constants are necessary to account for this effect!

A potential way to resolve these difficulties of the traditional Navier-Stokes model is to introduce the concept of the viscous *strength*; first introduced in Volokh (2009). In the present work, we enhance the mentioned model and, most importantly, *calibrate* it based on experimental data (Avila et al, 2011) of the onset of turbulence in pipe flow. *It is explicitly found that the maximal shear stress that can be sustained by water in the laminar flow is of the order of one Pascal*.

The paper is organized as follows. Section 2 presents the governing equations of the generalized Navier-Stokes model including a description of the viscosity failure. These governing equations are used in Section 3 to show the possibility of linear instability in the planar Couette flow between parallel plates. Section 4 explains why simple rheological measurements are not helpful in finding fluid strength. Then, the solution of the pipe flow problem is used in Section 5 to calibrate the viscous failure parameter based on the experimental results regarding the onset of turbulence. Section 6 presents a discussion of the proposed model.

## 2. Governing equations

The balance laws for mass and momentum in the absence of body forces take the following forms respectively:

$$\operatorname{div}\mathbf{v} = 0 , \tag{1}$$

$$\rho \frac{\partial \mathbf{v}}{\partial t} + \rho(\mathbf{v}\cdot\nabla)\mathbf{v} = \operatorname{div}\boldsymbol{\sigma} , \tag{2}$$

where $\rho$ is a constant mass density; $\mathbf{v}$ is a fluid particle velocity; $t$ is time; and $\boldsymbol{\sigma} = \boldsymbol{\sigma}^T$ is the Cauchy stress tensor.

The stress tensor is decomposed into two terms

$$\boldsymbol{\sigma} = -p\mathbf{1} + \boldsymbol{\tau} , \tag{3}$$

where $p$ is an unknown hydrostatic pressure; $\mathbf{1}$ is the second-order identity tensor; and $\boldsymbol{\tau}$ is the so-called viscous stress.

Traditionally, the constitutive model for the viscous stress in Newtonian fluids is set as follows

$$\boldsymbol{\tau} = 2\eta\mathbf{D} , \tag{4}$$

where $\eta$ is a *constant* viscosity coefficient and

$$\mathbf{D} = \frac{1}{2}(\nabla\mathbf{v} + \nabla\mathbf{v}^T) , \tag{5}$$

is the symmetric part of the velocity gradient.

Substituting (3)-(5) in (2) we get the Navier-Stokes equations, which should be completed with boundary/initial conditions for velocities in order to set an initial-boundary-value problem (IBVP).

Instead of the constant viscosity coefficient in (4), however, we will use the following constitutive model enforcing failure of viscous/frictional molecular bonds

$$\boldsymbol{\tau} = 2\eta * \mathbf{D} , \tag{6}$$

$$\eta * = \eta \exp[-(\mathbf{D}:\mathbf{D}/\phi^2)^m] , \tag{7}$$

where $\mathbf{D}:\mathbf{D} = D_{ij}D_{ij}$ in Cartesian coordinates and the sum is performed over the repeated indices; $m$ is a constant whose meaning is clarified below; and $\phi > 0$ is a *constant* defining the viscous fluid strength – Fig. 1.

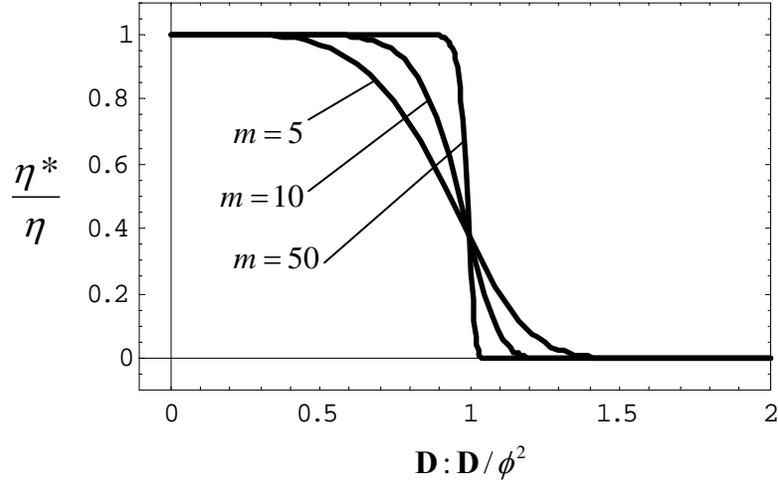

Fig. 1 The meaning of the viscosity coefficient in (7)

It is clear from Fig. 1 that the viscosity coefficient has two main modes

$$\eta^* = \eta \quad \text{when} \quad \mathbf{D} : \mathbf{D} < \phi^2 \, , \tag{8}$$

$$\eta^* = 0 \quad \text{when} \quad \mathbf{D} : \mathbf{D} > \phi^2 \, , \tag{9}$$

and by increasing $m$ it is possible to sharpen the step function.

The first mode, (8), corresponds to the classical Navier-Stokes viscosity with the full internal friction. The second mode, (9), corresponds to the loss of viscosity or internal friction. These two modes reflect upon Landau's remark that "…*for the large eddies which are the basis of any turbulent flow, the viscosity is unimportant*" (Landau and Lifshitz, 1987, Section 33: "Fully developed turbulence").

In the case of the shear flow, $\mathbf{D} = D_{12}(\mathbf{e}_1 \otimes \mathbf{e}_2 + \mathbf{e}_2 \otimes \mathbf{e}_1)$ and $\boldsymbol{\tau} = \tau_{12}(\mathbf{e}_1 \otimes \mathbf{e}_2 + \mathbf{e}_2 \otimes \mathbf{e}_1)$, the constitutive law, (6), takes the following dimensionless form

$$\frac{\tau_{12}}{\eta\phi} = 2\frac{D_{12}}{\phi}\exp(-2^m \frac{D_{12}^{2m}}{\phi^{2m}}) \, , \tag{10}$$

and it can be presented graphically – Fig. 2.

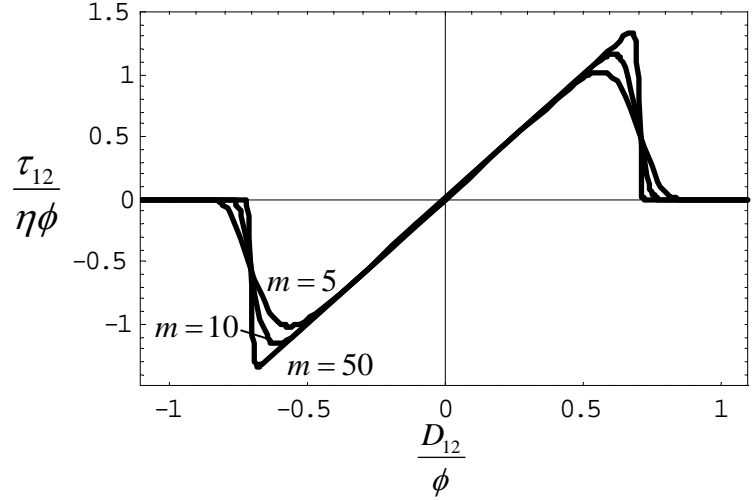

Fig. 2 Viscous stress versus deformation rate

The critical points of the viscosity failure in Fig. 2 have the following horizontal coordinates

$$\frac{D_{12}}{\phi} = \frac{\pm 1}{\sqrt{2}\sqrt[2m]{2m}} . \tag{11}$$

In the limit case of $m \to \infty$ we get the critical deformation rate

$$D_{12}^{cr} = \frac{\pm \phi}{\sqrt{2}}, \tag{12}$$

and the critical stress, which can be called the *viscous fluid strength*,

$$\tau_{12}^{cr} = \pm\sqrt{2}\eta\phi . \tag{13}$$

The traditional Navier-Stokes theory is obtained from the modified constitutive law when the fluid strength goes to infinity with $\phi \to \infty$.

It is worth noting that some viscosity thinning (Malek et al, 1996) takes place in the vicinity of the limit point when the fluid strength is approached. This phenomenon is negligible, being mostly a consequence of our analytical exponential expression for the constitutive law rather than a physical phenomenon.

*It is crucial to emphasize that the proposed model describes the breakdown of viscous bonds at the limit points in Fig. 2. Such limit points are absent in the traditional non-Newtonian or generalized Newtonian models of fluids. Thus, the proposed model should be interpreted as a Newtonian model with finite strength rather than a non-Newtonian model.*

We will complete the general modification of the Navier-Stokes theory by setting the linearized governing equations where the small perturbations of the motion dressed with tildes are superimposed on the existing motion. Varying equations (1)-(3) and (5)-(7) we get accordingly

$$\operatorname{div}\tilde{\mathbf{v}} = 0 \,, \tag{14}$$

$$\rho\frac{\partial\tilde{\mathbf{v}}}{\partial t} + \rho(\mathbf{v}\cdot\nabla)\tilde{\mathbf{v}} + \rho(\tilde{\mathbf{v}}\cdot\nabla)\mathbf{v} = \operatorname{div}\tilde{\boldsymbol{\sigma}} \,, \tag{15}$$

$$\tilde{\boldsymbol{\sigma}} = -\tilde{p}\mathbf{1} + \tilde{\boldsymbol{\tau}} \,, \tag{16}$$

$$\tilde{\mathbf{D}} = \frac{1}{2}(\nabla\tilde{\mathbf{v}} + \nabla\tilde{\mathbf{v}}^{T}) \,, \tag{17}$$

$$\tilde{\boldsymbol{\tau}} = 2\tilde{\eta}*\mathbf{D} + 2\eta*\tilde{\mathbf{D}} \,, \tag{18}$$

$$\tilde{\eta}* = -\frac{2\eta}{\phi^{2m}}(\mathbf{D}:\mathbf{D})^{m-1}(\mathbf{D}:\tilde{\mathbf{D}})\exp[-(\mathbf{D}:\mathbf{D}/\phi^{2})^{m}] \,. \tag{19}$$

The addition of the initial/boundary conditions of zero velocity perturbations completes the linearized IBVP.

## 3. Planar Couette flow

In this section we illustrate the generalized Navier-Stokes model developed in the previous section with the analysis of the planar Couette flow – see Drazin (2002) and Schmid and Henningson (2001) for review.

We assume that there is no pressure gradient and the velocity field has form: $\mathbf{v} = v_{1}(x_{2})\mathbf{e}_{1}$, where $\mathbf{e}_{1}$ is a unit base vector – Fig. 3.

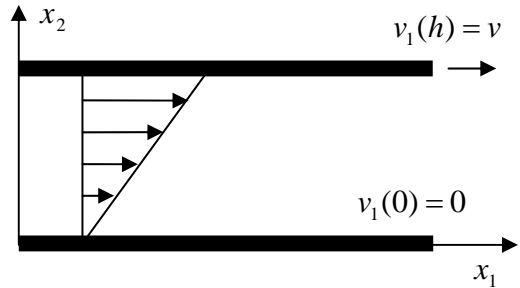

Fig. 3 Flow between parallel plates

In this case we have

$$\mathbf{D} = \frac{1}{2}\frac{\partial v_1}{\partial x_2}(\mathbf{e}_1 \otimes \mathbf{e}_2 + \mathbf{e}_2 \otimes \mathbf{e}_1)\,, \tag{20}$$

$$\mathbf{D} : \mathbf{D} = \frac{1}{2}\left(\frac{\partial v_1}{\partial x_2}\right)^2\,, \tag{21}$$

$$\boldsymbol{\tau} = \tau_{12}(\mathbf{e}_1 \otimes \mathbf{e}_2 + \mathbf{e}_2 \otimes \mathbf{e}_1)\,, \quad \tau_{12} = \eta\frac{\partial v_1}{\partial x_2}\exp\left[-\frac{1}{2^m\phi^{2m}}\left(\frac{\partial v_1}{\partial x_2}\right)^{2m}\right]\,, \tag{22}$$

and the reduced momentum balance (2) takes form

$$\frac{\partial \tau_{12}}{\partial x_2} = 0\,. \tag{23}$$

Substituting (22) in (23) and adding boundary conditions $v_1(0) = 0$ and $v_1(h) = v$ we find the following solution for velocity and stress fields

$$v_1 = v x_2 / h\,, \tag{24}$$

$$\sigma_{11} = \sigma_{22} = \sigma_{33} = -p; \quad \sigma_{12} = \sigma_{21} = \frac{\eta v}{h}\exp[-\frac{v^{2m}}{2^m h^{2m}\phi^{2m}}]\,. \tag{25}$$

Let us study the linear stability of the obtained solution. We assume that $\tilde{p} = 0$ and $\tilde{\mathbf{v}} = \tilde{v}_1(x_2)\mathbf{e}_1$. Then we have

$$\tilde{\mathbf{D}} = \frac{1}{2}\frac{\partial \tilde{v}_1}{\partial x_2}(\mathbf{e}_1 \otimes \mathbf{e}_2 + \mathbf{e}_2 \otimes \mathbf{e}_1)\,, \tag{26}$$

$$\tilde{\boldsymbol{\sigma}} = \tilde{\boldsymbol{\tau}} = \beta\frac{\partial \tilde{v}_1}{\partial x_2}(\mathbf{e}_1 \otimes \mathbf{e}_2 + \mathbf{e}_2 \otimes \mathbf{e}_1)\,, \tag{27}$$

$$\beta = \eta(1 - \frac{v^{2m}}{2^{m-1}h^{2m}\phi^{2m}})\exp[-\frac{v^{2m}}{2^m h^{2m}\phi^{2m}}]\,. \tag{28}$$

The momentum balance (15) reduces to

$$\rho\frac{\partial \tilde{v}_1}{\partial t} = \beta\frac{\partial^2 \tilde{v}_1}{\partial x_2^2}\,. \tag{29}$$

We further assume the following modes of the perturbed motion

$$\tilde{v}_1(x_2, t) = \text{constant} \cdot e^{\omega t}\sin(2\frac{\pi n}{h}x_2)\,, \quad n = 1, 2\ldots\,, \tag{30}$$

where boundary conditions are: $\tilde{v}_1(x_2 = 0, h) = 0$; and $\omega$ is a real constant.

Substituting (30) in (29) we find

$$\omega = -\frac{4\eta\pi^2 n^2 \beta}{\rho h^2}.$$  (31)

The Couette flow is stable when $\omega$ is negative and it loses stability when $\omega = 0$ and, consequently, $\beta = 0$. The latter condition gives the critical velocity of the top plate

$$v^{cr} = \frac{\sqrt{2}}{\sqrt[2m]{2}} h\phi.$$  (32)

and the critical Reynolds number

$$Re^{cr} = \frac{h\rho v^{cr}}{\eta} = \frac{\sqrt{2}}{\sqrt[2m]{2}} \frac{h^2 \rho \phi}{\eta}.$$  (33)

In the case of $m = 1$ we obtain the result which corrects the one corrupted by a numerical error in Volokh (2009).

For the limit case of the abrupt failure of viscosity where $m \to \infty$ we have

$$Re^{cr} = 1.41 \frac{h^2 \rho \phi}{\eta}.$$  (34)

It is interesting that in the case of the classical Navier-Stokes model where the fluid strength is infinite, $\phi \to \infty$, the flow is always stable with respect to the lateral perturbations – see also Romanov (1973), while in the case where strength is finite the flow can lose stability initiating the transition to turbulence.

### 4. Can viscous strength be calibrated in a viscometer/rheometer?

The material constants involved in the constitutive equations of Newtonian or non-Newtonian (Schowalter, 1978; Truesdell and Rajagopal, 2000) fluids are usually calibrated in simple rheological tests. Such a calibration is possible because the material behavior and the specific flow in the rheometer are *stable*. Thus, the stability of the process is a requisite for the model calibration.

Unfortunately, the calibration of the fluid strength is related to the onset of the material instability and the flow process becomes unstable. Thus, the classical rheological measurements are not completely adequate for the strength calibration. Indeed, let us assume that water is placed in a viscometer in the state of simple shear flow described in the previous section. In that case the breakdown of viscosity would be observed when all fluid particles arrive at the limit point in Fig. 2 *simultaneously*. That would be a measurable homogenous failure. However, the fluid material is never perfect and the viscous bonds do not fail simultaneously and homogeneously. The failure localizes. This localization is generally unpredictable - some viscous bonds endure while others break. The overall integral resistance of the flow will not change and the local bond failures will not be sensed by a viscometer. However, the local failure of viscosity can lead to the onset of the transition to turbulence. Therefore, the classical rheological tests are not good for the strength measurement, and experiments which track the onset of turbulence are better suited.

It is useful to set up an analogy between material failure in fluids and solids. Material failure in solids is a much better (yet not completely) understood phenomenon than material failure in fluids. Remarkably, the calibration of strength in solids displays similar difficulties to the calibration of strength in fluids. Indeed, a slab of concrete never fails homogeneously in biaxial tension tests where the stress-strain state is homogeneous – the failure localizes into cracks. It is impossible to predict where cracks should appear. It is only possible to find the critical stress at which cracks are expected to appear – the strength. Solids have shape, so the appearance of cracks is visible. Unfortunately, this is not the case for fluids. They have no shape, and so the localized failure of viscous bonds is invisible. However, the localized failure of viscous bonds can manifest itself in the onset of the transition to turbulence where viscosity is insignificant – see Landau's remark above.

## 5. Calibration of viscous strength: flow in a pipe

To track the onset of turbulence in order to find the viscous strength we consider the axisymmetric flow in a pipe – Fig. 4 – aiming at the calibration of the failure parameter: $\phi$.

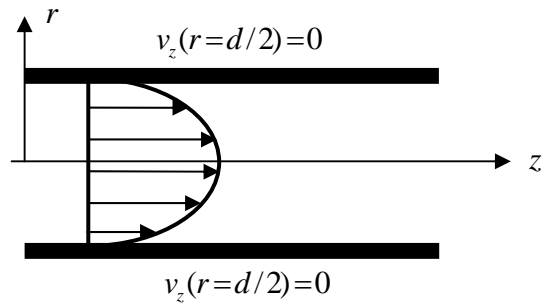

Fig. 4 Flow in a pipe

We use the cylindrical coordinates $(r, \varphi, z)$ with the unit base vectors

$$\mathbf{g}_r = \begin{pmatrix} \cos\varphi \\ \sin\varphi \\ 0 \end{pmatrix}, \quad \mathbf{g}_\varphi = \begin{pmatrix} -\sin\varphi \\ \cos\varphi \\ 0 \end{pmatrix}, \quad \mathbf{g}_z = \begin{pmatrix} 0 \\ 0 \\ 1 \end{pmatrix}. \tag{35}$$

We assume the axisymmetric distribution of velocities

$$\mathbf{v} = v_z(r)\mathbf{g}_z, \tag{36}$$

which obeys (1).

In this case we have

$$\mathbf{D} = \frac{1}{2}\frac{\partial v_z}{\partial r}(\mathbf{g}_r \otimes \mathbf{g}_z + \mathbf{g}_z \otimes \mathbf{g}_r), \tag{37}$$

$$\mathbf{D}:\mathbf{D} = \frac{1}{2}\left(\frac{\partial v_z}{\partial r}\right)^2, \tag{38}$$

$$\boldsymbol{\tau} = \tau_{rz}(\mathbf{g}_r \otimes \mathbf{g}_z + \mathbf{g}_z \otimes \mathbf{g}_r), \quad \tau_{rz} = \eta*\frac{\partial v_z}{\partial r}, \quad \eta* = \eta\exp\left[-\frac{1}{2^m\phi^{2m}}\left(\frac{\partial v_z}{\partial r}\right)^{2m}\right], \tag{39}$$

and the momentum balance (2) takes form

$$\frac{\partial}{r\partial r}\left\{r\eta*\frac{\partial v_z}{\partial r}\right\} + \alpha = 0, \tag{40}$$

$$\alpha = -\frac{\partial p}{\partial z}. \tag{41}$$

Differential equation (40) is transformed into a two-point boundary value problem by adding boundary conditions in the form

$$v(r = d/2) = 0, \quad \frac{\partial v_z}{\partial r}(r = 0) = 0. \tag{42}$$

When $(\partial v_z/\partial r)^2/2 < \phi^2$, we have $\eta* = \eta$ and we obtain the classical parabolic distribution of velocity by solving (40) and (42) analytically

$$v_z = \frac{d^2 - 4r^2}{16\eta}\alpha. \tag{43}$$

We assume now that the transition to turbulence takes place when the viscous bonds break and the solution of the flow problem (40)-(42) starts deviating from the parabolic law (43). The latter happens at the critical magnitude of the pressure gradient $\alpha^{cr}$ when the velocity gradient takes the form

$$\frac{\partial v_z}{\partial r} = -\frac{r}{2\eta}\alpha^{cr}, \tag{44}$$

and

$$(\mathbf{D}:\mathbf{D})^{cr} = \frac{1}{2}\left(\frac{r\alpha^{cr}}{2\eta}\right)^2. \tag{45}$$

The maximum on the right hand side of (45) is reached at the wall of the pipe

$$(\mathbf{D} : \mathbf{D})^{cr}_{\max} = \frac{1}{2}\left(\frac{d\alpha^{cr}}{4\eta}\right)^2. \tag{46}$$

The latter equation allows us to derive the strength parameter following (7) and Fig. 1

$$\phi = \sqrt{(\mathbf{D} : \mathbf{D})^{cr}_{\max}} = \frac{d\alpha^{cr}}{4\sqrt{2}\eta}. \tag{47}$$

It only remains to find the critical pressure gradient $\alpha^{cr}$ for the given diameter $d$ of the pipe.

At this point the experimental data is necessary. Usually, the Reynolds number,

$$Re = \frac{v_m \rho d}{\eta}, \tag{48}$$

is calculated based on the experimental data on the onset of turbulence under the flow with mean velocity $v_m$ and mass density $\rho$.

Formula (48) with the critical Reynolds number, $Re^{cr}$, of the onset of turbulence can be used to calculate the critical mean velocity

$$v_m^{cr} = \frac{\eta}{\rho d} Re^{cr}. \tag{49}$$

On the other hand, the critical mean velocity is related to the critical pressure gradient as follows

$$\alpha^{cr} = \frac{32\eta}{d^2} v_m^{cr}. \tag{50}$$

Substituting (49) in (50) we can find the critical pressure gradient in the form

$$\alpha^{cr} = \frac{32\eta^2}{\rho d^3} Re^{cr}. \tag{51}$$

Finally, substituting (51) in (47) we get

$$\phi = \frac{8\eta}{\sqrt{2}\rho d^2} Re^{cr}, \tag{52}$$

where for water we have

$$\rho = 10^3 (\text{kg/m}^3), \quad \eta = 10^{-3} (\text{N} \cdot \text{s/m}^2). \tag{53}$$

Avila et al (2011) reported that they observed transition to turbulence in a small pipe with the following parameters

$$d = 4 \cdot 10^{-3} (\text{m}), \quad Re^{cr} \approx 2400. \tag{54}$$

Substituting (53)-(54) in (52) we get

$$\phi = \frac{8 \cdot 10^{-3} \cdot 2400}{\sqrt{2} \cdot 10^3 \cdot 16 \cdot 10^{-6}} = 848.5 \, (\text{s}^{-1}), \qquad (55)$$

and the *viscous fluid strength* in (13) gets magnitude

$$\left| \tau_{12}^{cr} \right| = \sqrt{2} \cdot 10^{-3} \cdot 848.5 = 1.2 \, (\text{Pa}). \qquad (56)$$

Of course, the accuracy of the results given in (55) and (56) should not be overestimated because the precise evaluation of the critical Reynolds number is not a simple problem. Nevertheless, the order of the predicted viscous strength of water seems to be reasonable when compared, for example, to the strength of concrete ~$10^6$ Pa. It should not be overlooked in this comparison that stresses are triggered by different mechanisms in water and concrete. The latter does not matter. *Regardless of cause the maximal stress must be bounded in any material, fluid or solid!*

## 6. Discussion

In the present work we introduced a generalized Navier-Stokes constitutive model that allows for a description of the failure of viscous bonds through a parameter $\phi$, which is a material constant. This parameter means the maximum local 'equivalent' velocity gradient that can be reached before the viscous friction breaks. The introduction of the viscosity failure in the constitutive description of water was largely motivated by the desire to explain the subcritical transitions to turbulence based on material failure considerations. Particularly we showed based on the proposed model that shear flow between parallel plates can lose its stability under infinitesimal lateral perturbation in contrast to the classical Navier-Stokes considerations where such instability cannot occur in principle (Romanov, 1973).

It is crucial to emphasize that we not only proposed a new model with viscous strength but also calibrated it. For the latter purpose we considered flow in a pipe where the maximal velocity gradient takes place near the wall. We assumed that the onset of the flow instability observed in the experiments of Avila et al (2011) was triggered by the loss of viscosity near the wall and we found the corresponding magnitude of parameter $\phi$ – Eq. (55). Moreover, for the found parameter we estimated the maximal shear stress – fluid strength – that a layer of water can bear in a stable mode – Eq. (56). *This viscous strength of water is of the order of one Pascal*. The calibration of the model completes it for the purpose of future numerical simulations.

The proposed generalized Navier-Stokes model bounds the maximal stress that the viscous bond can sustain suppressing the pathological stress singularities that can appear in the traditional model.

The proposed generalized Navier-Stokes model also uncovers new prospects for the explanation of the remarkable phenomenon of the delay of the transition to turbulence due to the addition of a small amount of long polymer molecules in water. Indeed, the delay of the transition to turbulence can be qualitatively explained by the increase of the fluid strength due to the addition of the polymer. These polymer molecules create additional centers of attraction for water molecules. It can be assumed that polymers used in small amounts do not change the water viscosity as considerably as they change the water strength. Evolving parameter $\phi$ depending on the polymer concentration is a good candidate to describe the delay of the onset of the flow instability. It is very interesting to note that a similar situation is well known for solids where composite materials enjoy a higher overall strength compared to that of individual constituents.

Finally, we should note that the proposed mechanism of transition to turbulence through the material failure might not be the only mechanism and the transition to turbulence through the kinematic instability is not ruled out. The somewhat new ideas concerning the equations of hydrodynamics presented in this work should be further examined in the numerical simulations of various flows. Such broad studies could validate or pinpoint the applicability of the new model. We hope that the proposed calibrated model could attract the attention of the numerical analysts and encourage them to implement it in computer simulations.